\documentclass[12pt]{article}

\def\a{\alpha}

\def\b{\beta}

\def\th{\theta}

\def\s{\sigma}
\def\f{\phi}

\def\D{\Delta}
\def\e{\epsilon}

\def\d{\delta}
\def\m{\mu}
\def\n{\nu}
\def\l{\lambda}
\def\L{\Lambda}

\def\t{\tau}

\def\cd{{\cal D}}
\def\cc{{\cal C}}

\def\pr{\partial}
\def\to{\rightarrow}

\newcommand{\be}{\begin{equation}}
\newcommand{\ee}{\end{equation}}
\newcommand{\bea}{\begin{eqnarray}}
\newcommand{\eea}{\end{eqnarray}}

\begin{document}

\begin{titlepage}

\bigskip
\bigskip
\bigskip
\bigskip

\begin{center}

{\bf{\Large Spin Foam Models of Yang-Mills Theory Coupled to
Gravity }}

\end{center}
\bigskip
\begin{center}
 A. Mikovi\'c \footnote{E-mail address: amikovic@ulusofona.pt}
\end{center}
\begin{center}
Departamento de Matem\'atica e Ci\^{e}ncias de Computac\~ao,
Universidade Lusofona, Av. do Campo Grande, 376, 1749-024, Lisboa,
Portugal
\end{center}

\normalsize

\bigskip
\bigskip
\begin{center}
                        {\bf Abstract}
\end{center}
We construct a spin foam model of Yang-Mills theory coupled to
gravity by using a discretized path integral of the BF theory with
polynomial interactions and the Barret-Crane ansatz. In the
Euclidian gravity case we obtain a vertex amplitude which is
determined by a vertex operator acting on a simple spin network
function. The Euclidian gravity results can be straightforwardly
extended to the Lorentzian case, so that we propose a Lorentzian
spin foam model of Yang-Mills theory coupled to gravity.
\end{titlepage}
\newpage

\section{Introduction}

Spin foam (SF) models of quantum gravity represent a promising
approach for formulating a consistent theory of quantum gravity
\cite{bar,sfrev}. It is a discretized spacetime path-integral
approach, and the main result is the finitenness of the 3-geometry
to 3-geometry transition amplitude for a non-degenerate space-time
triangulation \cite{cpr}. Incorporating matter in the SF framework
has been started in \cite{msf}. However, the constructions given
there were algebraic, without a direct connection to the discrete
path-integral considerations. The difficulty in the path-integral
approach is that the usual matter fields couple to gravity via the
fierbein one-forms, while in the SF framework the fundamental
field is a 2-form $B$ which can be considered as an exterior
product of two fierbein one-forms. Still, in some cases the
coupling of matter can be expressed conveniently in terms of the
$B$ fields, like in the Yang-Mills (YM) theory case \cite{msf} \be
S_{YM} \propto \int \e_{abcd} B^{ab}\wedge B^{cd}\, Tr\, (F^{kl}
F_{kl}) \quad,\ee where $F_{kl} = e_k^\m e_l^\n F_{\m\n}$ and
$e_k^\m$ is the inverse of the fierbein matrix $e_\m^a$. $F$ is
the YM two-form field strength, and $Tr$ is the gauge group trace.

In such cases one can use the discretized path-integral techniques
developed for the $SU(2)$ BF theory by Freidel and Krasnov
\cite{fksf}, in order to derive the SF amplitude. Note that Oritti
and Pfeiffer have proposed recently a YM euclidian SF amplitude
without using the path-integral approach \cite{orpf}. In order to
better understand their result, we will study the discretized
path-integral for a general BF theory with polynomial in $B$
interactions by using the Friedel-Krasnov approach. The algebraic
structure of the corresponding SF amplitudes is such that it is
straightforward to adapt them via the Barrett-Crane ansatz
\cite{bce,bcl} to the case when the $B$ field is a simple
bivector, i.e. when it describes a realistic spacetime geometry.

In section two we generalize the Friedel-Krasnov results to the
case of an arbitrary Lie group $G$. In section three we perform
the Barrett-Crane reduction of the results from the section two.
In section four we construct the YM spin foam amplitudes. In
section five we present our conclusions.

\section{Path-integral approach}

Consider a BF theory with a polynomial interaction, depending only
on the $B$ field \be S = \int_M \langle B \wedge F \rangle +
\int_M P(B)\, d^4 x \quad,\ee where $M$ is a four-dimensional
manifold and $B$ is a two-form taking values in the Lie algebra
$\bf g$ of the Lie group $G$. $F =(dA_I + f_I^{JK} A_J\wedge A_K
)T^I$ is the curvature two-form taking values in $\bf g$ and
$\langle,\rangle$ is an invariant quadratic form on $\bf g$.

We would like to define the path integral \be Z = \int \,\cd A
\,\cd B \, e^{i S}\quad,\ee via the discretization procedure
provided by a simplical decomposition of $M$. We will use the
generating functional technique, so that we introduce \be Z[J] =
\int \,\cd A \,\cd B \, e^{i S + i \int_M \langle B\wedge J\rangle
}\ee and \be Z_0 [J] = \int \,\cd A \,\cd B \, e^{i \int_M \langle
B\wedge (F + J)\rangle}\quad.\ee Hence \be Z[J] = \exp \left( i
\int_M P \left(-i{\d\over\d J}\right)\right)Z_0 [J] \quad,\ee and
$Z = Z[0]$.

In order to define $Z_0$ let $\cc$ be a simplical complex
associated to $M$, and let $\cc^*$ be the dual complex. We denote
by $\s$, $\t$ and $\D$ a 4-simplex, a tetrahedron and a triangle
of $\cc$ respectively, and by $v$, $l$ and $f$ the corresponding
dual cells, i.e. a vertex, a link and a face, so that \be v = \s^*
\quad,\quad l = \tau^* \quad,\quad f = \Delta^* \quad.\ee Let $B$
field be piece-wise constant in the 4-polytopes formed by the
pairs $(\D,\D^* )$ such that the only non-zero components of $B$
lay in the $\D$ planes\footnote{This discretization procedure is
more suitable for treating interactions than the one used in
\cite{fksf}, which was based on the delta-function B fields. See
\cite{dav} for basic notions about the forms on simplical
complexes.}. Then \be S = \sum_{\D}\langle B_\D , F_f \rangle +
\sum_{\D,...,{\D}^{\prime}} C_{I\cdots
I^\prime}(\D,...,{\D}^{\prime} )B^I_\D \cdots
B^{I^\prime}_{{\D}^{\prime}} \quad,\ee where \be B_\D = \int_\D B
\quad,\quad F_f = \int_f F \label{simpa}\quad.\ee

Given the variables (\ref{simpa}), we define \be Z_0 [J]= \int
\prod_{l} dA_l \prod_{\D} dB_\D e^{i\sum_{\D}\langle B_\D ,F_f +
J_f \rangle} \quad,\ee where $A_l = \int_l A$. Following the ref.
\cite{fksf} we will fix the measures of integration $dA$ and $dB$
such that \be Z_0 [J] = \int \prod_l dg_l \prod_f \d
\left(\prod_{l\in f}\left(g_{l}e^{J_f^{(v)} }\right)\right)
\quad,\label{gtz}\ee where $g_l = e^{A_l}$, and $J_f^{(v)}$ are
the Lie algebra elements associated to the vertices $v$ of the
face $f$ such that \be \sum_{v \in f}J_f^{(v)}= J_f \quad.\ee The
formula (\ref{gtz}) differs from the one given in \cite{fksf} by
the absence of the factor $\prod_{v\in f}P(J_f^{(v)})$. We do not
put this factor because the simpler expression (\ref{gtz}) is also
invariant under the gauge transformations \be g_l \to h_v \,g_l \,
h_{v^\prime}^{-1} \quad,\quad J_f (v) \to h_v \,J_f (v) \,
h_{v}^{-1}\quad,\quad h_v ,h_{v^\prime}\in G \,, \ee where $v$ and
$v^\prime$ are the ends of the link $l$ \cite{fksf}. Also when $J
\to 0$ the expression (\ref{gtz}) gives the standard formula for
the partition function for the BF theory \cite{o,bar}. Hence we
take the expression (\ref{gtz}) as the definition of $Z_0(J)$.

The expression (\ref{gtz}) can be further simplified by taking \be
J_f^{(v)} = J_f^{(v^\prime )}= \cdots = {J_f \over n_f}\quad,\ee
where $n_f$ is the number of vertices of the face $f$. By
rescaling $J_f \to n_f J_f$ we can get rid of the $n_f$ factors.
By using the group theory formulas \be \d (g) = \sum_{\L}
\textrm{dim}\,\L \,\chi_{\L} (g)\quad, \ee and \be \int_G \,dg
\,D^{(\L_1)}_{\a_1\b_1} (g) D^{(\L_2)}_{\a_2\b_2}
(g)D^{(\L_3)}_{\a_3\b_3} (g) D^{(\L_4)}_{\a_4\b_4} (g) =
\sum_{\iota} C^{{\L_1}{\L_2}{\L_3}{\L_4}(\iota)}_{\a_1 \a_2 \a_3
\a_4} \left(C^{{\L_1}{\L_2}{\L_3}{\L_4}(\iota)}_{\b_1 \b_2 \b_3
\b_4}\right)^*\quad,\ee where $D^{(\L)}$ is the group
representation matrix in a representation $\L$, and $\chi_\L$ is
the corresponding trace, as well as the associated graphical
calculus \cite{fksf,bar}, one arrives at \be Z_0 = \sum_{\L_f
,\iota_l }\prod_f \textrm{ dim}\,\L_f \prod_l C(\iota_l )\prod_v
\Psi_5 (\L_{f(v)},\iota_{l(v)},e^{J_{f(v)}})\quad, \label{pent}\ee
where $\Psi_5$ is the spin network function associated to the
pentagram graph and $C(\iota)$ is a constant coming out from the
recoupling theory calculus.

The problem with (\ref{pent}) is that the sum over the irreps
diverges. This sum can be regularized in the $J=0$ case by passing
to the quantum group $G_q = U_q (\bf g)$, $q$ root of unity
\cite{cky}. One could generalize this procedure for the non-zero
$J$ case, but since in this paper we are not interesting in the
topological gravity case, we will not elaborate upon this.

\section{The Barrett-Crane ansatz}

General relativity can be understood as a BF theory with
constraints since \be \int \sqrt{|g|}\,R \, d^4 x = \int
\e^{abcd}e_a \wedge e_b \wedge R_{cd} = \int B^{ab}\wedge
R_{ab}\quad,\ee where $e_a$ are the fierbein one-forms and
$R_{ab}$ is the spin-connection curvature two-form. Hence $G =
SO(4)$ in the Euclidian case or $G=SO(3,1)$ in the Lorentzian case
and the constraint is $B^{ab}= *(e^a \wedge e^b)$. In the SF
framework this constraint is implemented as a restriction on the
set of irreps $\{\L_f \}$ such that \be \langle T(\L_f ), *T(\L_f
)\rangle = \e^{abcd}T_{ab}(\L_f ) T_{cd}(\L_f ) = 0 \quad,\ee
where $T(\L)$ are the generators of $G$ in the representation $\L$
\cite{bce,bb,bcl}. The corresponding solutions are called simple
irreps, which we denote as $N$. Equivalently, $N$ is the irrep
containing an $SO(3)$ invariant vector \cite{fksi}.

Given the set of simple irreps $\{N_f \}$ one postulates that $Z$
is of the same form as in the topological case \be Z = \sum_{N_f}
\prod_{f} A_2 (N_f ) \prod_{l} A_{1} (N_{f(l)})\prod_{v}A_{0}
(N_{f(v)}) \quad,\label{sfa}\ee but now the amplitudes $A_i$ are
different. In the Euclidian case $N=(j,j)$, $j\in \frac12\bf Z$,
and the face amplitude $A_2 = \textrm{dim}\, N = (2j+1)^2$
\cite{bce}. In the Lorentzian case $A_2 = p^2 dp$, where
$N=(0,p)$, $p\ge 0$ \cite{bcl}. The vertex amplitude $A_0$, which
is associated to the pentagram spin net, is given now as \be A_0
(N_1 , \cdots, N_{10}) = \int_{G/H} \prod_{v=1}^5 dx_v
\prod_{l=1}^{10} K_{N_l}(x_{v(l)}, x_{v^{\prime}(l)})\quad,\ee
where $x_v$ are the points in the homogeneous space $G/H$ and
$H=SO(3)$ \cite{bce,bss,fksi,bcl}. The propagators $K$ are given
as \be K_N (x,y) = \langle 0| D^{(N)}(g_x g_y^{-1})|0\rangle
\quad,\label{prop}\ee where $|0\rangle$ is the invariant vector
from $N$, which corresponds to the identity irrep of the subgroup
$H$ contained in $N$. There is a freedom in choosing the edge
amplitude $A_1$, and a choice which makes the sum (\ref{sfa})
finite is the spin net amplitude for the theta-four graph
\cite{per,cpr}.

Given these results and the formula (\ref{pent}) for the
topological case, it is natural to propose the following
expression for $Z_0 (J)$ in the non-topological case \be Z_0 (J) =
\sum_{N_f} \prod_{f} A_2 (N_f ) \prod_{l} A_{1}
(N_{f(l)})\prod_{v}\Phi_{5} (N_{f(v)};e^{J_{f(v)}})
\,,\label{sfj}\ee where \be \Phi_5 (N_1 , \cdots, N_{10};
e^{J_1},\cdots, e^{J_{10}}) = \int_{G/H} \prod_{v=1}^5 dx_v
\prod_{l=1}^{10} K_{N_l}(x_{v(l)},
x_{v^{\prime}(l)};e^{J_l})\,,\label{snf}\ee and the propagator \be
K_N (x,y;e^J) = \langle 0| D^{(N)}(g_x \, e^J\, g_y^{-1})|0\rangle
\quad,\ee is the "source" propagator introduced in \cite{fksi}.
Hence $\Phi_5$ is an example of the simple spin network functions
introduced in \cite{fksi}.

\section{The Yang-Mills model}

The YM action on $\cc$ is given by \cite{ren} \be \l^2 S_{YM}=
\sum_{\D}{A(\D^* ) \over A(\D)}\left( Re\,Tr\,\tilde U(\D) - n
\right) \quad, \ee where $\tilde U$ is the triangle holonomy
associated to the gauge group $\tilde G$ and $n$ is the dimension
of the matrix $\tilde U$. $A(\D)$ is the area of the triangle
$\D$, while $A(\D^*)$ is the area of its dual face. Let us rewrite
this action as \be \l^2 S_{YM}=\sum_{\D}{6V(\D,\D^* )\over
A^2(\D)}\left( Re\,Tr\,\tilde U(\D) - n \right)\quad,\ee where
$V(\D,\D^* )=\frac16 A(\D)A(\D^*)$ is the 4-volume of the
4-polytope $(\D,\D^* )$ \cite{dav}. This can be further rewritten
as \be \l^2 S_{YM}={1\over 3}\sum_\s \sum_{\D \in\s}{6V(\D,\D^*
)\over A^2(\D)}\left( Re\,Tr\,\tilde U(\D) - n \right)\quad,\ee
which is suitable for the spin foam formalism.

The Oriti-Pfeiffer proposal \cite{orpf} can be understood as an
approximation \be V(\D,\D^* ) \approx \sum_{\D^\prime \in\s
}C(\D,\D^\prime ) \langle B_{\D},*B_{\D^\prime}\rangle \quad.\ee
This is an approximation because $V(\D,\D^*)$ depends in general
on the $B$ fields from the adjoint 4-simplices which share the
polytope $(\D,\D^*)$, so that for an exact formula one would have
to include the terms with $\D^\prime \in \s^\prime$. A further
approximation is to take the $C$'s to be constant up to the
orientation sign factors $\textrm{sign}(\D,\D^\prime)$. This
happens for the symmetric lattices when $V(\D,\D^*) = (2/5)
V(\s)$, where $V(\s)$ is the 4-volume of the 4-simplex $\s$. Since
\be V(\s ) ={1\over30\cdot4!}
\sum_{\D,\D^\prime}\,\textrm{sign}\,(\D,\D^\prime ) \langle
B_{\D},*B_{\D^\prime}\rangle \quad,\ee one obtains \be
C(\D,\D^\prime ) \approx {4\over
30\cdot5!}\,\textrm{sign}\,(\D,\D^\prime ) \quad.\ee

One can then define \be Z_{YM} = \int \prod_\e d\tilde g_\e
\prod_l dg_l \prod_{\D} dB_\D \, e^{iS_0 + iS_{YM}}= \int \prod_\e
\,d\tilde g_\e \,\tilde Z_{YM} \quad,\ee where $\e$'s are the
edges of the complex $\cc$. By using the previous results we
obtain\be \tilde Z_{YM} \approx\sum_{N_f} \prod_{f} A_2 (N_f )
\prod_{l} A_{1} (N_{f(l)})\prod_{v}\left[e^{-i\b{\hat
S}_{v}}\Phi_{5} (N_{f(v)};e^{J_{f(v)}})\right]_{J=0}
\,,\label{yma}\ee where $\b$ is a numerical constant and \be {\hat
S}_{v} = \sum_{\D,\D^\prime \in\s}{\left( Re\,Tr\,\tilde U (\D) -
n \right) \over A^2(\D)}\, \textrm{sign}\,(\D,\D^\prime
)\left\langle{\pr\over \pr J_f}\,,\,*{\pr\over \pr
J_{f^\prime}}\right\rangle\quad.\label{ymv}\ee

A natural further simplification is to take \be
A(\D)=\sqrt{j(j+1)} \quad,\ee in the Euclidian case \cite{orpf}.
In the Lorentzian case the labels of the triangles can be related
to the areas of space-like triangles \cite{bcl}, and hence it is
natural to take \be A(\D)= \sqrt{p^2 +1} \quad.\ee

One can show  that \be \left[e^{-i\b{\hat S}_{v}}\Phi_{5}
(N_{f(v)};e^{J_{f(v)}})\right]_{J=0} = \sum_{\a,\b}C_{\a_1 \cdots
\a_{10}}^{\b_1 \cdots \b_{10}}\langle {\a_1 \cdots \a_{10}}|
e^{-i\b{\hat S}_{v} (T)}|{\b_1 \cdots
\b_{10}}\rangle\quad,\label{trf}\ee where $\hat S (T)$ is the
operator obtained from (\ref{ymv}) by replacing the $\pr J$'s with
the $T(N)$'s. It acts in the space $H_v = \bigotimes_{\D\in \s}
V(N_f )$. The coeficients $C$ are given as products of five
Barrett-Crane intertwiners \be S^{N_1 \cdots N_4}_{\a_1 \cdots
\a_{4}} = \sum_N C^{N_1 \cdots N_4 (N)}_{\a_1 \cdots
\a_{4}}\quad.\ee

From (\ref{trf}) it is clear that in the Euclidian case the
expression $e^{-i\b\hat S_v}\Phi_5 |_{J=0}$ is well defined since
$N$'s are finite-dimensional and hence all the sums are finite.
The expectation value in (\ref{trf}) is the matrix element of an
exponential of a finite matrix, which exists.

In the Lorentzian case, the operator representation (\ref{trf}) is
not that useful for showing that $e^{-i\b \hat S_v}\Phi_5 |_{J=0}$
is well-defined. We expect it to be well defined because in the
Euclidian case the propagators\footnote{The extra factor
$(2j+1)^{-1}$ follows from the formula (\ref{prop}), since the
$SU(2)$ invariant vector has a normalisation factor
$(2j+1)^{-1/2}$.} \be K_j (x,y) = {\sin (2j +1) d(x,y)\over (2j+1)
\sin d(x,y) }\quad,\quad \cos d (x,y) = x\cdot y \quad,\ee get
deformed when the sorce $J$ is turned on as \be \cos d \to \cos d
\,\cos |J| + \sin_J d\,\sin |J| \quad,\label{dfr}\ee where \be
 \sin_J d = \cos \th \sin \f \cos\f_J + \sin\th \cos\f \cos\th_J\ee
for $x=(\cos\th,\sin\th,0,0)$ and $y=(\cos\f,\sin\f \cos\a,\sin\f
\sin\a,0)$. The angles $\th_J$ and $\f_J$ are determined by the
3-vectors $x$ and $J$ and $y$ and $J$, respectively, which are
associated to the corresponding $SU(2)$ subgroup elements. The
deformation (\ref{dfr}) is qualitatively similar to the
deformation $d \to d + |J|$, so that it is no surprise that the
integral (\ref{snf}) is finite, given that the $J=0$ integral is
finite. Since the Lorentzian propagator \be K_p (x,y) = {\sin p\,
d(x,y)\over p\, \sinh d(x,y) }\quad,\quad \cosh d(x,y) = x\cdot y
\quad, \ee can be obtained from the Euclidian one via analytical
continuation $2j + 1 \to ip$ and $d \to id$, we then expect to
have a deformation \be \cosh d \to \cosh d \,\cos |J| + \sinh_J
d\,\sin |J| \quad,\label{ldfr}\ee where \be
 \sinh_J d = \cosh \th \sinh \f \cosh\f_J + \sinh\th \cosh\f \cosh\th_J\ee
for $x=(\cosh\th,\sinh\th,0,0)$ and $y=(\cosh\f,\sinh\f
\cos\a,\sinh\f \sin\a,0)$. This is similar to the deformation $d
\to d + i|J|$. Since the $J=0$ integral (\ref{snf}) converges, we
then expect that the $J\ne 0$ integral will converge as well.

\section{Conclusions}

The discretized path-integral approach to spin foam amplitudes
based on the generating functional technique is very useful,
because the algebraic structure of the expressions one obtains in
the topological case is such that it can be straightforwardly
extended to the non-topological cases. This algebraic structure is
the tensor category of irreducible representations of the symmetry
group, so that the spin foam transition amplitudes can be
understood as functors constructed from the special morphisms,
which are the spin networks \cite{mct}. Even in the
non-topological case, when the simple irreps $N$ do not form a
tensor category, one is actually dealing with the tensor category
of the irreps of the subgroup $H$ \cite{mct}, which in the YM case
are just the trivial identity irreps, so that it is natural that
we have proposed the generating functional in terms of the simple
spin network functions.

Because of this, we consider our proposal for the vertex amplitude
(\ref{trf}) more appropriate for the spin foam formalism than the
one proposed in \cite{orpf}, which is \be {Tr_{H_v} (e^{-i\b\hat
S_v (T)})\over \textrm{dim}\,H_v }= \sum_{\a}(\textrm{dim}\, H_v
)^{-1}\langle {\a_1 \cdots \a_{10}}| e^{-i\b{\hat S}_{v}
(T)}|{\a_1 \cdots \a_{10}}\rangle\quad.\ee Although this is a
reasonable proposal, this expression does not have a simple
interpretation in terms of spin network evaluations. Consequently,
it is not clear what would be a Lorentzian extension of this
expression. On the other hand, our expression (\ref{trf}) is based
on a simple spin network evaluation, and hence it makes sense in
the Lorentzian case.

We expect that the YM amplitude (\ref{yma}) is finite both in the
Euclidian and in the Lorentzian case because of the very rapid
convergence properties of the $J=0$ sums in the Euclidian case
\cite{bac}.

Note that the structure of the vertex amplitude is such that one
can use the perturbative expansion for small $\b$\be \left[e^{-i\b
\hat S_v }\Phi_5 \right]_{J=0} = \left[(1 - i\b \hat S_v + \cdots
)\Phi_5 \right]_{J=0} = A_0 -i\b \left[\hat S_v \Phi_5
\right]_{J=0} + \cdots\quad. \label{pea}\ee The second term in
(\ref{pea}) can be interpreted as the pentagram spin network
amplitude with the insertions of two endomorphisms $T(N)$ and
$T(N^\prime)$ at the edges $N$ and $N^\prime$. If we neglect the
higher-order terms in the expansion (\ref{pea}) we will obtain a
spin foam amplitude which is a sum of amplitudes where a string of
pentagram spin networks is connected by the matter edges carrying
the adjoint representation of $G$. This amplitude is of the
general type proposed in \cite{msf}, but it is more complicated
then the ones considered there. In \cite{msf}, only the amplitudes
where the matter edges link the vertices of the pentagrams were
considered, while the amplitude corresponding to (\ref{pea}) has
the matter edges linking the edges of the pentagrams and there are
$T(N)$ insertions. This demonstrates that in order to understand
completely the coupling of matter in the spin foam formalism, it
is necessary to study the corresponding discretized path integral.

\end{document}